\documentclass[sigconf]{acmart}
\usepackage{xcolor}
\usepackage{tabularx}

\newcolumntype{Y}{>{\centering\arraybackslash}X}

\definecolor{RevisionColor}{HTML}{000000}

\newcommand\revision[1]{\textcolor{RevisionColor}{#1}}
\usepackage{xcolor}
\usepackage{makecell}
\usepackage{array}
\usepackage{placeins}

\AtBeginDocument{%
  }

\copyrightyear{2026}
\acmYear{2026}
\setcopyright{cc}
\setcctype{by-nc-nd}
\acmConference[CHI '26]{Proceedings of the 2026 CHI Conference on Human Factors in Computing Systems}{April 13--17, 2026}{Barcelona, Spain}
\acmBooktitle{Proceedings of the 2026 CHI Conference on Human Factors in Computing Systems (CHI '26), April 13--17, 2026, Barcelona, Spain}
\acmPrice{}
\acmDOI{10.1145/3772318.3791128}
\acmISBN{979-8-4007-2278-3/2026/04}

\begin{document}

\title[Revisiting Worker-Centered Design: Tensions, Blind Spots, and Action Spaces]{Revisiting Worker-Centered Design: Tensions, Blind Spots, and Action Spaces}


\author{Shuhao Ma}
\email{shuhao.ma@tecnico.ulisboa.pt}
\orcid{0000-0003-3639-3064}
\affiliation{%
  \department{ITI / LARSyS, Instituto Superior Técnico}
  \institution{University of Lisbon}
  \city{Lisbon}
  \country{Portugal}
}

\author{John Zimmerman}
\email{johnz@cs.cmu.edu}
\orcid{0000-0001-5299-8157}
\affiliation{%
    \department{Human-Computer Interaction Institute}
    \institution{Carnegie Mellon University}
    \city{Pittsburgh}
    \state{Pennsylvania}
    \country{USA}
}

\author{Valentina Nisi}
\orcid{0000-0002-8051-3230}
\email{valentina.nisi@tecnico.ulisboa.pt}
\affiliation{%
  \department{ITI / LARSyS, Instituto Superior Técnico}
  \institution{University of Lisbon}
  \city{Lisbon}
  \country{Portugal}
}

\author{Nuno Jardim Nunes}
\email{nunojnunes@tecnico.ulisboa.pt}
\orcid{0000-0002-2498-0643}
\affiliation{%
  \department{ITI / LARSyS, Instituto Superior Técnico}
  \institution{University of Lisbon}
  \city{Lisbon}
  \country{Portugal}
}

\renewcommand{\shortauthors}{Ma et al.}
\begin{abstract}
Worker-Centered Design (WCD) has gained prominence over the past decade, offering researchers and practitioners ways to engage worker agency and support collective actions for workers.
Yet few studies have systematically revisited WCD itself, examining its implementations, challenges, and practical impact. 
Through a four-lens analytical framework that examines multiple facets of WCD within food delivery industry, we identify critical tensions and blind spots from a Multi-Laborer System perspective. 
Our analysis reveals conflicts across labor chains, distorted implementations of WCD, designers’ sometimes limited political-economic understanding, and workers as active agents of change. 
These insights further inform a Diagnostic-Generative pathway that helps to address recurring risks, including labor conflicts and institutional reframing, while cultivating designers’ policy and economic imagination. 
Following the design criticism tradition, and through a four-lens reflexive analysis, this study expands the action space for WCD and strengthens its relevance to real-world practice.
\end{abstract}



\maketitle

\section{Introduction}
Worker-Centered Design (WCD) has emerged as a prominent topic in HCI and design~\cite{fox_workercenter_dis20, Kim_success}, particularly in studies of vulnerable workers, including food couriers~\cite{shuhao_centered_23, Shaikh_chi24}, drivers~\cite{gloss_designing_2016, ning_ma_stakeholder18}, service workers~\cite{Koo_serviceworker}, bus operators~\cite{Akridge_bus}, warehouse workers~\cite{Cheon_warehouse}, caregivers~\cite{Lustig_carework}, sex workers~\cite{Hamilton23} and content moderators~\cite{Lee_contentmoderator}.
WCD reorients the design focus from user communities to worker communities. It leverages empathic~\cite{shuhao_centered_23}, participatory~\cite{Angie_stakeholders23}, and speculative design approaches~\cite{kirman_thinking_2022} to elevate workers' voices throughout the design process. These developments reflect a growing recognition of worker agency in WCD and highlight the role of HCI and design in supporting collective action.

WCD's emergence creates the need for systematic reflection that revisits WCD literature and situated research practice. Critique has long played an important role in HCI and design research, helping the field grow and improve~\cite{Pierce_criticality, jz_critique}. 
Dominant design methods like user-centered design have been critically examined for either unintentionally or intentionally obscuring structural inequalities, reinforcing power asymmetries, and universalizing particular user experiences~\cite{Norman_harmful, Steen2012, Steen-ucdtension, Buchanan1992, Bardzell_feminist10, costanza-chock_design_2020}. 
Similarly, participatory design, which originated in workplace contexts, has also been critically examined, particularly for how unequal power relations during participatory processes may compromise its democratic aims~\cite{Kensing1998-kd, Vines_conpd, Harrington_19, Mackay2023-iz}. Critiques of HCI and design methods have produced important methodological advances and valuable reframings.

We propose that WCD research may overlook key methodological tensions and systemic conflicts found in user-centered design~\cite{Steen-ucdtension}, participatory design~\cite{Vines_conpd}, service design~\cite{Van_Amstel2025}, and broader design traditions, thereby risking the reproduction of long-standing design shortcomings.
This problem is urgent given persistent global labor inequality and the rapid expansion of platform economies~\cite{Schor2017, TAYLOR2023106234}. As WCD research increasingly influences design practice, understanding its limitations becomes essential to ensure that interventions genuinely support worker welfare rather than inadvertently perpetuating harm. The stakes are particularly high in service sectors such as ride-hailing and food delivery, where algorithmic management~\cite{lee_algorithmics_15}, labor precarity~\cite{Qadri_infras, vandaele_vulnerable_2022}, and policy interventions~\cite{Angie_policy} continuously reshape working conditions.

\revision{Drawing from our four-year engagement with gig workers~\cite{shuhao_centered_23, shuhao_decide_24, Ma_chi25, ma_mapping_2023, shuhao_dis_dc}}, including food couriers and drivers, we are prompted to ask three provocative questions: How can WCD avoid reproducing design shortcomings? Who counts as the ``worker''? Why is WCD difficult to realize in practice?
Informed by our WCD projects, along with prior studies and real-world cases, we apply a four-lens analytical framework to examine design-labor relationships, workers' lived experiences, platform design processes, and broader political-economic contexts.

Our analysis identifies critical blind spots in current WCD research and its connection to its intended impact in practice.
We introduce a Multi-Laborer System perspective that reveals conflicts across labor chains, distorted worker-centered implementations, and designers' limited capacity to engage with policy and economic forces that shape intervention viability. Meanwhile, we recognize spaces where workers' own agency plays a central role in reshaping their environments, even without direct design intervention.
Our paper contributes to WCD research by reflecting on design risks that are seldom considered through the perspective of labor conflict.
Then, we propose a Diagnostic–Generative Pathway for designing for workers, intended as a reference to address labor conflict, institutional reframing, and political-economic dynamics while equipping designers with actionable means to create interventions.

Our reflections are not general critiques or dismissals of prior WCD work. Instead, we follow \citet{Pierce_criticality} in viewing design criticism as provisional, open-ended, and incomplete, emphasizing the importance of ongoing dialogue and iteration.
Our revisit of WCD research surfaces blind spots and offers new directions for future inquiry, with the aim of broadening its actionable space so that its intended impact in practice can be more fully realized.

We begin by reviewing the key paradigms of WCD, then critique and reorient HCI and design methods in relation to it. Next, we introduce three provocative questions, present our method and findings on Multi-Laborer Systems, and conclude with a Diagnostic–Generative Pathway that highlights implications for expanding actionable design spaces. Finally, we discuss the significance of this work for HCI and design communities, its applications in practice and education, and its limitations, inviting researchers, practitioners, and educators to engage further with this domain.

\section{Related Work} \label{relatedwork}

This section presents the outcomes of WCD and examines how broader HCI and design methods have been critiqued and reoriented to support our revisit of WCD. These reviews form the basis for the three provocative questions we raise.

\subsection{Worker-Centered Design}\label{relatedwork_WCD}
\begin{figure*}
  \centering
  \includegraphics[width=\textwidth]{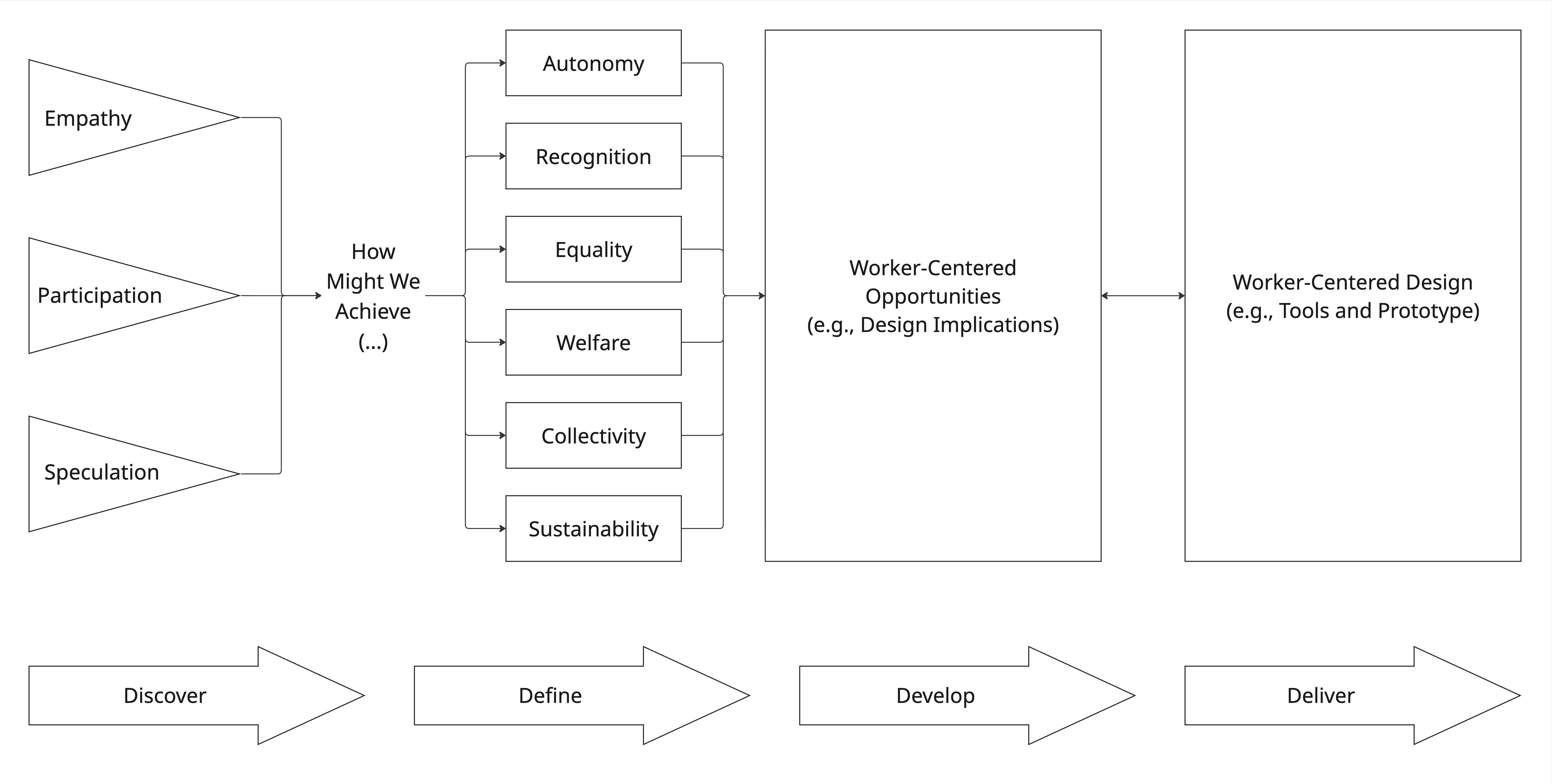}
  \caption{Paradigm and Outcomes of Worker-Centered Design -- We summarized worker-centered research in HCI and design, mapping its paradigms and outcomes onto the stages of a typical design process. The figure illustrates how different methods and contributions align with each stage: in the \textbf{\textit{DISCOVER}} phase, three primary approaches are employed -- empathy, participation, and speculation; in the \textbf{\textit{DEFINE}} phase, studies generate interpretations of workers' needs at multiple levels; in the solution phase, research \textbf{\textit{DEVELOP}} design opportunities and \textbf{\textit{DELIVER}} prototypes and real products.}
  \Description{A process diagram showing the paradigm and outcomes of WCD across four design stages: Discover, Define, Develop, and Deliver. On the left, three triangular blocks labeled Empathy, Participation, and Speculation feed into the question ``How Might We Achieve (…)'', representing the Discover phase. This leads to the Define phase, where six thematic areas are listed vertically: Autonomy, Recognition, Equality, Welfare, Collectivity, and Sustainability. These themes connect to the next phase labeled ``Worker-Centered Opportunities (e.g., Design Implications)'', which in turn links to ``Worker-Centered Design (e.g., Tools and Prototype)'' on the right, representing Develop and Deliver phases. Arrows indicate a linear and iterative flow through these stages.}
  \label{workercentereddesign}
\end{figure*}

Over the past decade, HCI and design have shown a growing interest in WCD, evident in venues such as CHI~\cite{fox_workercenter_dis20}, CSCW~\cite{Ming_cscw, Tang_cscwlabor}, DIS~\cite{Pillai_discare, shuhao_centered_23}, and CHIWORK~\cite{chiwork_22, Kim_chiwork, Munoz_chiwork}. 
This body of work places greater emphasis on the needs of worker communities. To trace this trajectory, we reviewed relevant research across these venues and mapped it onto a design process, examining how worker needs are discovered, defined, and translated into WCD outcomes (see Figure \ref{workercentereddesign}), illustrating how WCD contributes at different stages of a design process~\cite{Zimmerman07}.

\textbf{\textit{DISCOVER} Worker Needs.} The HCI and design community has integrated worker perspectives into the discovery stage of the design process. 
Diverse methods for identifying worker needs have meaningfully supported the agency of workers involved in WCD, including:

\begin{itemize}
    \item Empathy: Interviews and ethnographic fieldwork to understand workers' everyday experiences and challenges~\cite{Hernandez_chi24, Doggett_farmworkers, Kusk25}. Such studies provide a holistic understanding of workers and a foundation for developing solutions.
    \item Participation: Co-design and participatory methods that enable workers to participate directly in the design process~\cite{do_sousveillance, hsieh_co-designing_2023}. Such studies foster a more democratic design and research process by emphasizing workers' participation in generating potential solutions.
    \item Speculation: Speculative and future-oriented methods that invite workers to imagine alternative scenarios~\cite{alvarez_speculative_freelancer22, Lallemand_fiction}. Speculative methods are often combined with participatory approaches, extending the focus to possible futures rather than existing solutions.
\end{itemize}

Empathy, speculative, and participatory approaches all aim to center workers' lived experiences to better define their needs. To the best of our knowledge, WCD research consistently centers workers in design inquiry, amplifies their voices, grounds solutions, and opens new ways of engaging with labor conditions and well-being.

\textbf{\textit{DEFINE} Worker Needs.} Through the above approaches, WCD researchers have identified critical areas where workers' needs intersect with sociotechnical, economic, and political systems. We synthesize these insights into six key themes.

\begin{itemize}
    \item Autonomy: The extent to which workers can control their schedules, tasks, and work practices~\cite{yao_together_2021, lee_algorithmics_15, Hernandez_chi24, xiao2025labor_algorithm}.
    \item Recognition: Ensuring that invisible forms of labor are acknowledged and treated with dignity~\cite{irani_turkopticon, Spektor_discarded, xiao_invisible, Akridge_24}.
    \item Equality: Guaranteeing equity in pay, gender, race, and access to opportunities~\cite{Ming_wagetheft,Ning_22, Koo_serviceworker}.
    \item Welfare: Safeguarding workers' physical health, mental well-being, and social protections~\cite{So_techwork_chi, Lee_contentmoderator}.
    \item Collectivity: Strengthening solidarity, cooperation, and collective action among workers~\cite{Salehi_dynamo, irani_turkopticon}.
    \item Sustainability: Securing the long-term viability of work while addressing its social impacts~\cite{Kim_success, Ma_chi25}.
\end{itemize}

These themes reflect significant progress by HCI and design researchers in identifying diverse needs among diverse worker groups. Building on these insights, WCD research has revealed both complexities and opportunities for intervention. These needs are closely interconnected. For example, how autonomy affects mental health and how recognition shapes equity. Such entanglements indicate important challenges for researchers seeking to develop effective and holistic design responses.

\textbf{\textit{DEVELOP}~Design~Opportunities}~\textbf{\&}~\textbf{\textit{DELIVER}~Solutions}. \revision{As we~\cite{ma_mapping_2023} observed} in the analysis of design contributions in gig worker research, most studies, after identifying worker needs, proceed to the \textit{\textbf{DEVELOP}} stage by generating design opportunities (e.g.~\cite{shuhao_centered_23, Hernandez_chi24}). 
As noted in \textbf{\textit{DISCOVER} Worker Needs}, this stage is often grounded in worker engagement. As a result, many contributions are constructive yet remain largely conceptual, offering suggestions for platform features as well as broader insights for policy, organizational practices, and future research. 
Importantly, many design opportunities focus on platform design, reflecting WCD's expectations for future platform features, services, algorithms, and policies.
A key issue is the lack of frameworks in WCD to validate or implement design opportunities, creating a gap between expected implications and their practical adoption~\cite{shuhao_decide_24}. 

Some studies go further by delivering prototypes~\cite{Kasunic_tales, Qiu_avatar}, which we see as a form of constructive design~\cite{Koskinen_lab_field}. These exploratory prototypes play a crucial role in bridging the gap between conceptual insights and fully realized tools, offering tangible steps toward real-world application.
For instance, data-sharing platforms have been prototyped to explore how workers might resist platforms or influence policy~\cite{Hsieh_Gig2Gether}. 
While not implemented at scale, these interventions enrich our understanding of design possibilities and help chart potential paths forward for WCD.

There are also rare but notable cases where prototypes successfully moved beyond the lab to create real-world impact, most prominently in early research on Amazon Mechanical Turk. 
Some projects advanced into the \textit{\textbf{DELIVER}} stage by introducing tools that directly supported workers, such as Turkopticon~\cite{irani_turkopticon, irani_designtroubles} and Dynamo~\cite{Salehi_dynamo}. Turkopticon, in particular, stands out as a pioneering intervention that significantly empowered collective action among workers~\cite{irani_turkopticon}. Such examples highlight how WCD holds the capacity to effect meaningful change beyond academic settings by translating research into tangible support for worker agency.

Yet, we noted that most WCD contributions remain at the stage of informing design rather than producing immediate changes in practice~\cite{ma_mapping_2023}, and their long-term impact is often difficult to track once the research concludes.
Next, we examine the challenges WCD faces through the lens of diverse design methods. This reflection seeks to constructively revisit prior work and strengthen future directions for designing with and for workers.

\subsection{Critiquing and Reorienting HCI and Design Methods for WCD}

Critique has long played an important role in HCI and design research, helping the field grow and improve~\cite{Pierce_criticality, jz_critique}. This section examines key critiques and transitions in design to situate our revisit of WCD. These reflections are closely tied to how design methods can be operationalized in WCD, but they are not intended as general critiques of prior WCD work. Instead, we are aligned with \citet{Pierce_criticality}'s framing of design criticism as provisional, open-ended, and incomplete, highlighting the value of continued dialogue and iteration.

The first concept related to WCD is human-centered design or user-centered design (UCD). 
UCD has been successful in improving usability and meeting user needs, yet it has faced sustained critique for falling short in addressing wicked problems that require systemic, long-term, and collaborative approaches~\cite{Steen2012, Steen-ucdtension, Buchanan1992, Norman2015}. 
Meanwhile, UCD often fails to include marginalized populations.
HCI and design researchers are increasingly highlighting the tendency of UCD to privilege the individual user while overlooking broader social structures and power relationships~\cite{Bardzell_feminist10, costanza-chock_design_2020}. \citet{Bardzell_feminist10} argues that the neutral user assumed in traditional design often defaults to male experiences, thereby marginalizing women and racial minorities, and emphasizes the importance of including diverse users in the design process.

These sustained critiques show that even well-intentioned approaches can cause harm by overlooking structural injustice, reinforcing power imbalances, or flattening diverse experiences into a single, neutral user~\cite{Bardzell_feminist10, costanza-chock_design_2020, Steen-ucdtension}. 
Unlike traditional UCD, which often aligns with capital interest, WCD prioritizes principles such as care and equity, aiming to support workers' well-being, dignity, and agency within broader socio-technical systems.
Like UCD, WCD must prioritize its central subject, which in this case is the worker, while also engaging with other stakeholders and addressing complex, often contentious challenges. 
We argue that this intersection presents the first provocative question for the future of WCD: How can WCD avoid reproducing design shortcomings?

A second key concept related to WCD is participatory design (PD), which has its roots in efforts to democratize the workplace~\cite{Erling_pdc10, Björgvinsson01062012}. As discussed in Section \ref{relatedwork_WCD}, PD methods have been widely adopted in WCD research, giving rise to a number of worker-led initiatives such as Turkopticon~\cite{irani_turkopticon, irani_designtroubles} and Dynamo~\cite{Salehi_dynamo}. More recently, scholars like \citet{lee_algorithmics_15} and \citet{Angie_centerwork} have built on this tradition by co-designing algorithms and data probes with gig workers, further advancing the commitment to worker participation in platform design. These efforts reflect a growing recognition of worker agency in WCD and highlight the role of HCI and design in supporting collective action.
However, we want to carefully highlight the subtle gap between research through PD and the origins of PD itself. Early PD, in Scandinavia, focused on real-world collaboration among workers, unions, and stakeholders to co-create solutions~\cite{muller_kuhn_pd93, Bjerknes_1987}. Over time, PD shifted from democratizing the workplace to democratizing innovation~\cite{Erling_pdc10}. 
Most research through PD efforts involved workers in the design process but rarely led to real-world change beyond the research setting, often due to short-term academic deployments or the absence of broader stakeholder involvement~\cite{ma_mapping_2023, shuhao_decide_24}.
This is not intended as a critique, but as an acknowledgment of the complexity inherent in PD.

On the technical side, implementing PD requires care, as prior studies have raised cautious perspectives on the politics of design and user participation~\cite{Kensing1998-kd}. Some argue that users may not know or may be unwilling to articulate their needs~\cite{Vines_conpd}, while others note that PD activities are shaped by privilege and bias~\cite{Harrington_19}. Still others warn that placing too much emphasis on what users say can erode designers' responsibility, vision, and creativity, which are essential to the design process~\cite{Mackay2023-iz}.
On the ecological side, aligning WCD with broader concerns such as designer responsibility, capitalist structures, and consumer expectations remains a major challenge~\cite{Vallas2020-dj}. These issues extend beyond HCI and design, as seen in the global rise of cooperatives~\cite{Bunders2022-jn}. Despite their good intentions and long history, many cooperatives continue to struggle with sustainability, governance, funding, and power dynamics, and most ultimately fail~\cite{Kramper2012-ly, Garrido2007-me}. WCD faces similar barriers, which remain unresolved and require further attention. The complexity of worker participation and agency raises another provocative question: Why is WCD so difficult to realize in practice?

Another concept closely related to WCD is service design. Service design has been adopted to address some of the limitations of user-centered design by promoting more holistic approaches to thinking and designing~\cite{roto_sdux_21}. Service design tools, such as stakeholder mapping, and design principles, such as value co-creation, strongly resonate with the aims of WCD~\cite{Stickdorn18}. It is important to note that workers are not absent from service design practices. For instance, service blueprints often include both frontstage and backstage staff roles.

However, compared to UCD and PD, research that applies service design to understand workers' experiences remains relatively limited. An overlooked opportunity lies in using service design thinking to explore value co-creation among workers, consumers, and platforms~\cite{shuhao_centered_23}. Recent work by \citet{penin_bloomsbury_2026} underscores the inseparable relationship between service design, service workers, and the wider service industry, opening space for worker-centered service design. Yet in sectors such as food delivery and ride-hailing, which rely on extensive stakeholders, WCD research often focuses on a single worker role rather than the broader network of labor involved. From a service design perspective, this raises a provocative question: Who counts as the ``worker'' in WCD?

In addition to UCD, PD, and SD, our community has increasingly turned to approaches such as community-led design~\cite{costanza-chock_design_2020}, justice-oriented design~\cite{Dombrowski_SJOD}, and value-sensitive design~\cite{Friedman2019-VSD}. For example, \citet{Friedman2019-VSD} extend user-centered design by moving beyond immediate usability concerns to explicitly incorporate human values, such as equity, privacy, and justice, into the design process for both direct and indirect stakeholders. These approaches build on and respond to critiques of earlier design approaches, and each relates to WCD to varying degrees.

We analyzed the outcomes of WCD and introduced design critiques to better understand how WCD research connects to its intended impact in practice. Outside academia, its real-world influence remains uncertain. For example, current design practices are often neither user-centered~\cite{Green2021-cu} nor worker-centered~\cite{shuhao_decide_24}, as both consumers and workers are treated as resources for extraction. Drawing from broader critiques and transitions across UCD, PD, SD, and related approaches, we identified three provocative questions.  
\section{Three Provocative Questions for WCD}
\label{questions}

Our review shows that worker agency is increasingly recognized, and collective action is supported in HCI and design. However, many contributions remain limited to lab contexts and rarely scale to broader worker communities. Positioning WCD within the wider design research landscape reveals that it may still replicate persistent shortcomings found in traditional methods. Our reflections are grounded in both our design projects and prior research, aiming to unpack the relationship between WCD research and its intended real-world impact. Our reflection does not diminish the value of WCD, as such efforts are essential in the face of ongoing global labor inequality and exploitation. Instead, we emphasize the need to better connect HCI and design research with real-world impact and create space for improving future work.

\begin{itemize}

\item \textbf{How can WCD avoid reproducing design shortcomings?}
Even with the best intentions, WCD research can slip into familiar pitfalls of other design methods. Can WCD become harmful? How can we advance worker well-being without reinforcing these dynamics?

\item \textbf{Who counts as a ``worker'' in WCD?}
Research often centers on a single group, such as drivers, couriers, content creators, and caregivers, yet these workers exist within larger labor systems of interdependence and hidden work. When we say worker, whose labor are we really centering, and whose remains invisible?

\item \textbf{Why is WCD so difficult to realize in the real world?}
HCI and design research labs often function as Utopian spaces. Without real-world adoption, the effectiveness of design opportunities and solutions cannot be tested.
Many WCD projects remain confined to labs, with few transferring to practice. What prevents these design solutions from scaling?

\end{itemize}

\section{Method}\label{method}

We selected the app-based food delivery industry as the focal domain for this study due to our prolonged engagement in this area since 2021. Food delivery platforms have attracted intense scrutiny in recent years due to their impact on gig workers and society~\cite{kusk_lr22, ma_mapping_2023}. 
A wealth of HCI and design research has explored the daily work experiences of food couriers~\cite{shuhao_centered_23, vandaele_vulnerable_2022, Chen_silent, Shaikh_chi24, Hernandez_chi24} and a few prototyped systems to support their practices~\cite{Yoon20, Ding22}. Yet, existing research has tended to focus less on industry or ecosystem perspectives and has paid only limited attention to broader impacts associated with technology beyond the couriers themselves~\cite{shuhao_decide_24}.

Food delivery is a contested site of platform labor, concentrating critical HCI concerns such as algorithmic management~\cite{lee_algorithmics_15, Chen_AlgorithmChina}, labor precarity~\cite{Kusk25}, urban infrastructure~\cite{Qadri_infras}, and bias~\cite{Ning_22}. Ongoing regulatory and market shifts make the industry a dynamic testing ground for tensions between design, technology, governance, and worker welfare~\cite{Ma_chi25}, offering a timely context for rethinking how WCD can address systemic challenges.
Building on prior work (see Section \ref{relatedwork}) and \revision{our own research published at CHI~\cite{Ma_chi25}, DIS~\cite{shuhao_centered_23, shuhao_decide_24, shuhao_dis_dc}, and IASDR~\cite{ma_mapping_2023},} we propose a four-lens analytical framework that examines multiple facets of the food delivery industry.
By combining complementary lenses, the framework enriches interpretation, offsets the blind spots, and supports the reflexive analysis of WCD projects. 
Our approach aligns with multimethodological traditions that emphasize the representation of real-world complexity and the guidance of interventions in multiple stages~\cite{Mingers1997}, while also resonating with emerging lens-based practices in HCI that use multiple interpretive frameworks to produce actionable insights~\cite{Williams2014, Yang_WWH}. 
In particular, we draw on \citet{Yang_WWH}, who integrated prior research and their own studies to explain why Human-AI interaction is uniquely difficult to design, offering methodological guidance for our work.

\subsection{Understanding the Food Delivery Industry through Four Analytical Lenses}

To explore these provocative questions about WCD, our analysis seeks to capture the breadth of the food delivery industry. We approach this through four analytical lenses: foundation (design and labor relations), frontstage (workers' lived experiences), backstage (platform design processes), and offstage (political and economic contexts). 
These four lenses capture the conditions under which WCD unfolds. Our reflexive analysis~\cite{virginia_ref_19} can strengthen the field by revealing tensions and blind spots and by examining when and how WCD can, or cannot, take shape in a holistic context.

\begin{itemize}
\item
\textbf{Foundation - Design and Labor Relationship:} We begin from the recognition that design interventions are never neutral but always shape labor dynamics~\cite{gloss_designing_2016, crawford_atlas_2021}. \revision{Our review~\cite{ma_mapping_2023} and labor design work~\cite{Ma_chi25} identified and developed this lens, showing that design choices across products, interactions, services, and infrastructures profoundly shape working conditions.}
For example, task allocation algorithms shape the autonomy of couriers~\cite{yao_together_2021, lee_algorithmics_15}, while urban planning that ignores mobility workers creates further barriers~\cite{shuhao_centered_23}. This foundational lens highlights how WCD practices reconfigure labor relations among couriers and stakeholders, with cascading effects on work and well-being.

\item
\textbf{Frontstage - Workers' Lived Experiences:} Our second lens focuses on the lived realities of food couriers. \revision{Our empirical studies in Europe~\cite{shuhao_centered_23} and China~\cite{Ma_chi25},} complemented by global research, reveal recurring challenges across contexts: intense pressure, social isolation, and limited long-term prospects~\cite{kusk_lr22, Kusk25, kirman_thinking_2022, Chen_silent, Shaikh_chi24, vandaele_vulnerable_2022}. At the same time, social, political, and economic differences shape local variations in the way platform design issues manifest~\cite{Wood2019}. This lens highlights the consequences of platform design by situating everyday struggles of workers within broader socioeconomic, sociopolitical, and sociotechnical conditions, aligning with critical computing and design justice in foregrounding nuanced experiences of marginalized groups~\cite{costanza-chock_design_2020, Irani_postcolonial}.

\item
\textbf{Backstage - Platform Design Process:} \revision{This lens is crucial because our other review paper~\cite{ma_mapping_2023} shows that most WCD efforts focus on informing platform design.}
The third lens examines how platforms are built and governed. Building on HCI research that highlights designers, managers, operators, and developers as stakeholders in co-designing worker-centered systems~\cite{Angie_stakeholders23, hsieh_co-designing_2023, Tuco_21, richardson_22}, \revision{our prior work engaged designers from eight platforms across five countries~\cite{shuhao_decide_24}.} These accounts reveal how corporate priorities and resource constraints shape design outcomes. 

\item
\textbf{Offstage - Political and Economic Contexts:} The final lens examines offstage forces such as policy and market dynamics. Although less visible in platform interactions, policymakers and capital actors strongly shape industry trajectories and worker welfare~\cite{yang_policy_24chi, Whitney2015, heskett_creating_2009}. \revision{Our work has engaged this lens across most of our publications~\cite{shuhao_centered_23, shuhao_decide_24, Ma_chi25} as well as ongoing projects.}
Regulations such as the General Data Protection Regulation~\cite{GDPR16} and the European Accessibility Act~\cite{EUR_EAA19} impose transparency and accessibility requirements that directly influence platform design~\cite{shuhao_decide_24}.
Market shifts, such as the 2025 Price War among three major food delivery platforms (JD, Eleme, Meituan) in China, rapidly redefined worker benefits and business conditions~\cite{pricewar_chinadaily1, pricewar_chinadaily2, pricewar_economist, pricewar_reuter, pricewar_wallstreet}.
This lens highlights how political and economic forces shape the conditions for WCD, creating both opportunities and risks. Following \citet{Spektor_discarded}, we incorporated policy changes, industry reports, and public media to compare these dynamics with insights from the other lenses.

\end{itemize}

\subsection{Data Analysis}

\begin{figure*}
  \centering
  \includegraphics[width=\textwidth]{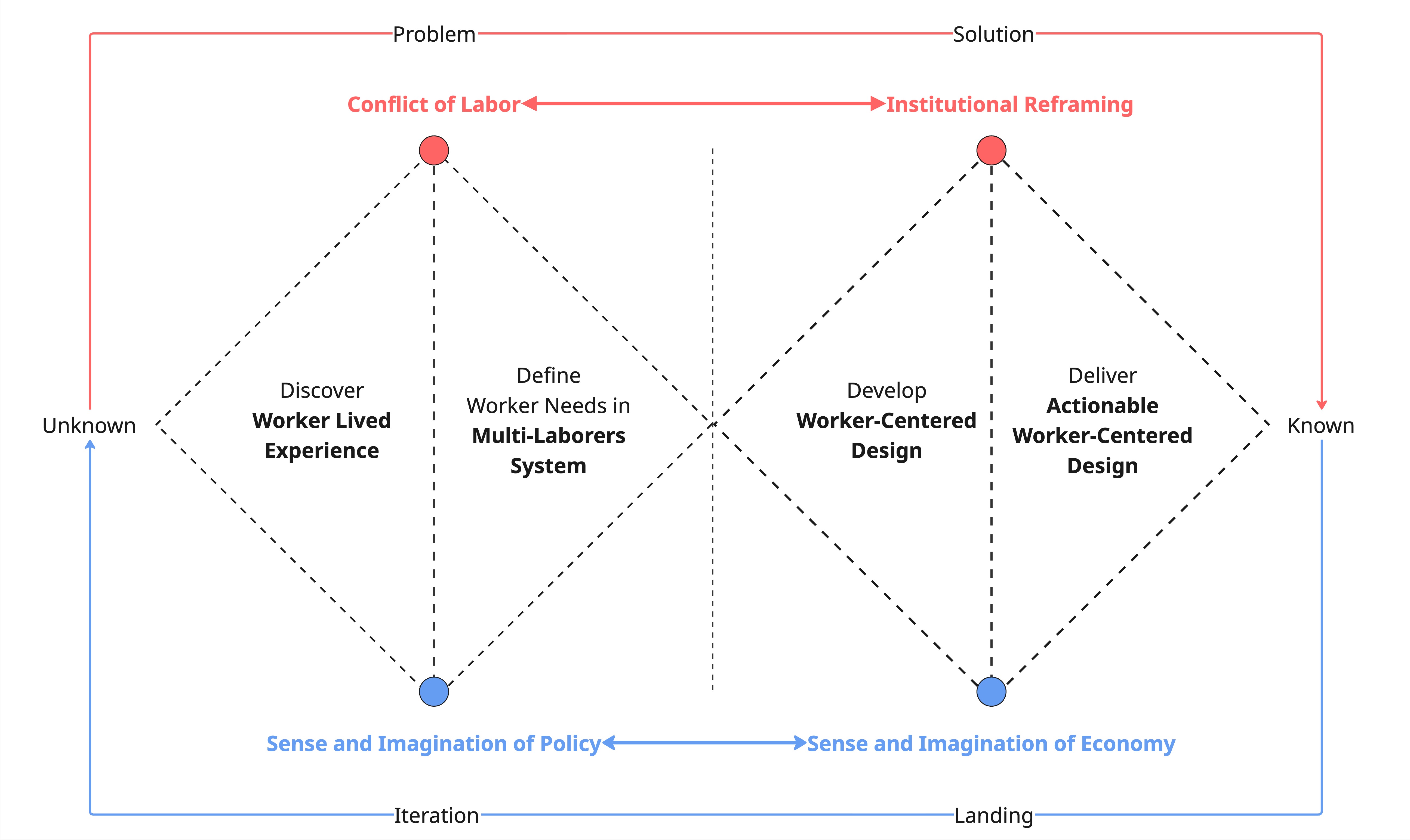}
  \caption{A Diagnostic–Generative Pathway -- Based on our analysis of Multi-Laborer Systems, we propose a Diagnostic–Generative Pathway for design in relation to workers. This figure illustrates how the Pathway can be applied within the double diamond model. The red diagnostic path spans the top of the diagram, starting with ``Conflict of Labor'' above the first diamond and ending with ``institutional reframing'' above the second. The blue generative path runs along the bottom, beginning with ``Sense and Imagination of Economy'' and ending with ``Sense and Imagination of Policy'' below the second.}
  \Description{A conceptual diagram presenting a Diagnostic–Generative Pathway overlaid on the double diamond model. The process is divided into four phases: Discover, Define, Develop, and Deliver. Two diamonds formed by dashed lines represent divergent and convergent phases in problem and solution spaces. The red diagnostic path spans the top of the diagram, starting with ``Conflict of Labor'' above the first diamond and ending with ``institutional reframing'' above the second. This path reflects critical tensions encountered during the discovery and definition of worker needs. The blue generative path runs along the bottom, beginning with ``Sense and Imagination of Economy'' and ending with ``Sense and Imagination of Policy'' below the second. This path represents potential directions for iteration and landing during the development and delivery of actionable WCD. The overall flow moves from ``Unknown'' on the left to ``Known'' on the right, emphasizing the transformation of problem understanding into implementable solutions. The figure highlights how diagnostic and generative processes intersect with WCD.}
  \label{revisedworkercentered}
\end{figure*}

We conducted a reflexive analysis~\cite{virginia_ref_19} drawing on literature, empirical data, and public media resources, structured around four lenses: foundation, frontstage, backstage, and offstage, while continually cross-checking insights and debating three provocative questions.
The process was informed by service design. It captures stakeholder involvement, value co-creation, and service blueprinting~\cite{Stickdorn18}.
Our analysis revealed recurring patterns and led to two outcomes. 
First, in Section \ref{multi-laborer}, we identified blind spots in existing approaches, illustrated through a Multi-Laborer System (see Figure \ref{multilaborer}) that prior research had overlooked.
Second, in Section \ref{pathway}, we introduce a Diagnostic–Generative Pathway for designing in relation to workers. 
We use the double diamond model (Discover, Define, Develop, Deliver) as an illustrative example (see Figure \ref{revisedworkercentered}) because its four stages are commonly used to represent how design knowledge can be contributed over different design stages, as needed~\cite{Zimmerman07}. This does not suggest that WCD work has followed or should follow this specific process.
By situating WCD within specific stages of the model, we highlight where it is most crucial to diagnose blind spots that hinder the addressing of worker needs. 

\subsection{Positionality}

We are designers and researchers from interaction design, service design, and participatory design backgrounds, with extensive experience researching labor issues such as gig work, migrant labor, and care work. This collective expertise shapes a service-oriented perspective, viewing workers and platforms as part of broader socio-technical systems, rather than as isolated user interfaces. Service design emphasizes integrated thinking across stakeholder touchpoints and organizational levels~\cite{Stickdorn18}, enabling a holistic approach to WCD that considers its role within platform ecosystems and labor markets. Our aim is not to reject WCD, but to foster critical reflection and development. Our approach is constructive: building on the strengths of WCD while interrogating its limitations to push the practice toward greater ethical and social impact.
As our analysis builds on prior work, we acknowledge its limitations and subjectivities. We engaged in critical dialogue, questioning why some WCD interventions fail to gain traction and whose interests are prioritized in design choices. These reflections revealed differing interpretations, while our framework remains a work in progress shaped by specific experiences rather than an exhaustive account.
\section{Multi-Laborer System}

\label{multi-laborer}

\begin{figure*}
  \centering
  \includegraphics[width=\textwidth]{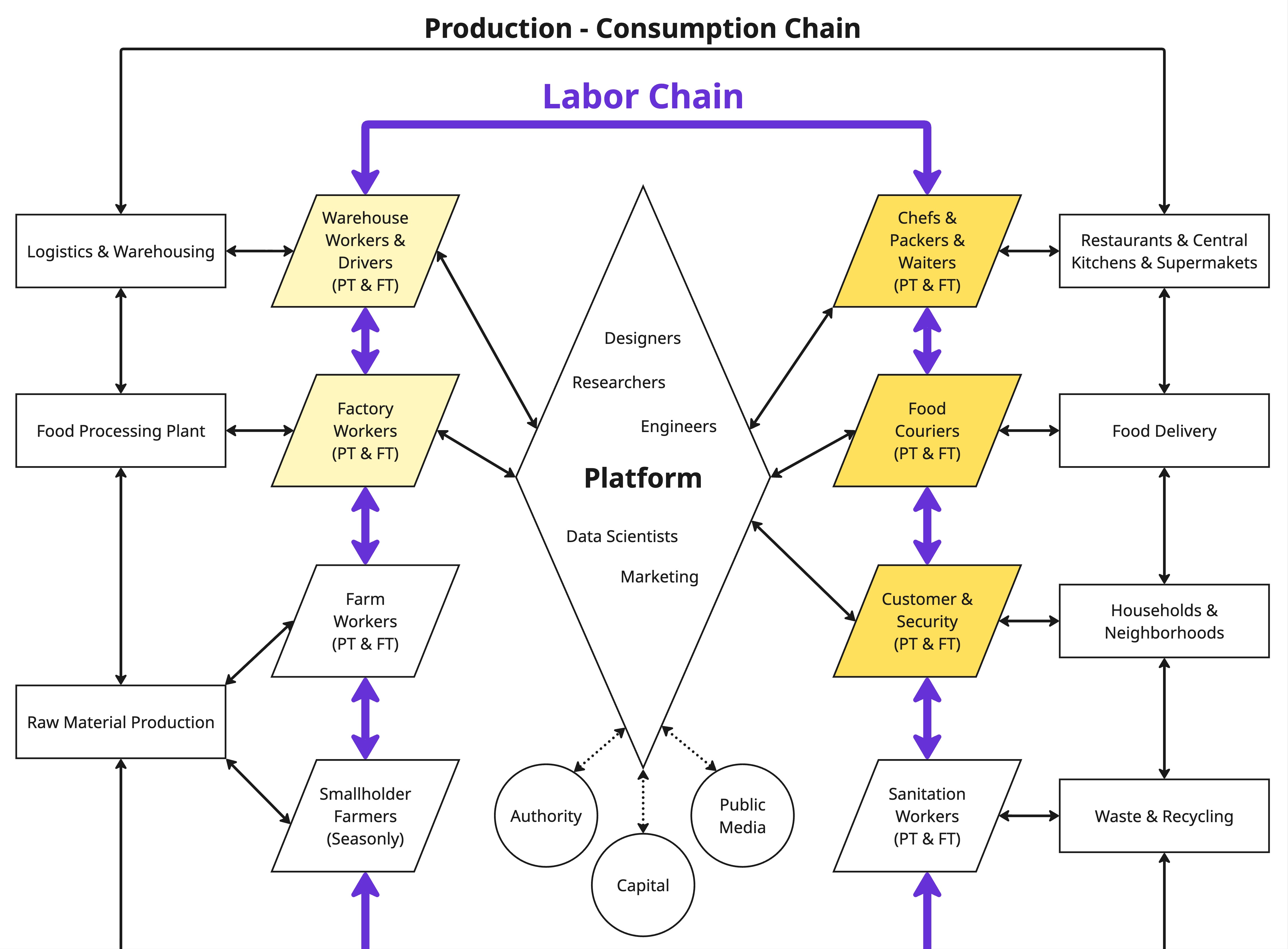}
  \caption{Multi-Laborer System -- This diagram conceptualizes the food delivery industry as a Multi-Laborer System, supported by various groups of workers involved throughout the production-to-consumption chain, operating under employment arrangements ranging from part-time (PT) to full-time (FT). The purple line highlights the ``Labor Chain''. At the center of the system is the platform, which not only coordinates couriers but also shapes the operations of restaurants, warehouses, and upstream food processing. Peripheral actors -- including authorities, capital investors, and public media -- interact with the platform, collectively influencing labor relations and the broader industry landscape.}
  \Description{A systems diagram representing the food delivery industry as a Multi-Laborer System along a production–consumption chain. A thick purple line labeled ``Labor Chain'' traces a path through multiple worker groups, from smallholder farmers to sanitation workers. These worker groups include farm workers, factory workers, warehouse workers and drivers, chefs and packers and waiters, food couriers, customer and security, and sanitation workers, all labeled as PT (part-time) and FT (full-time). At the center of the diagram is a large diamond-shaped node labeled ``Platform,'' encompassing roles such as designers, researchers, engineers, data scientists, and marketing. The platform connects to all worker groups, as well as to external actors shown as circles—Authority, Capital, and Public Media. The platform also influences logistics, food processing, restaurants, and delivery flows on both sides of the diagram. Arrows illustrate the flow of goods and services across the system and the influence of the platform over various entities. The visual emphasizes the central role of the platform in coordinating labor and shaping industry practices.}
  \label{multilaborer}
\end{figure*}

When we step back from a single worker-centered view, what comes into focus is a vast, platform-mediated socio-technical system of food. This system reconfigures the entire chain of production and consumption, from farming, processing, and logistics to dining, delivery, and waste (see Figure \ref{multilaborer}).
Food delivery is not new, but design, technology, capital, and innovation have reshaped it at every stage, altering both labor and everyday life.
For example, rising delivery volumes reduce in-restaurant dining, pushing restaurants toward smaller scales; the growth of pre-prepared meals has created new central-kitchen roles that displace traditional chefs (see Figure \ref{photoE} in Appendix); and the expansion of courier workforces has reshaped labor markets and domestic routines.

We propose the Multi-Laborer System to visualize the diverse labor roles and hidden tensions within service platforms.
Figure \ref{multilaborer} maps the Multi-Laborer System, placing the platform at its center as it mediates among multiple actors (such as workers, restaurants, factories, logistics providers, and regulators), each operating under distinct labor arrangements.
What emerges are not only relations of value co-creation, familiar from service design, but also frictions and conflicts, as highlighted in prior critiques of platform labor. While platforms promise value for these actors, they also redistribute risks, obscure dependencies, and create tensions across the chain.

We understand tension as encompassing both value and conflict, and treat these as potential entry points for design~\cite{Tatar_designtension, Kozlovski2022}. 
Beyond platform coordination of production and consumption, we emphasize overlooked opportunities for design to engage with labor relations, inequities, and the systemic transformations that food delivery sets in motion.
In what follows, we present our analysis of the tensions between design and labor in the food delivery industry:
i)~conflicts among laborers;
ii)~how these conflicts, shaped by capital and other power dynamics, lead to distorted forms of WCD;
iii)~how the pace of political and economic change outstrips designers' familiarity with it; and
iv)~how, beyond a single Multi-Laborer System, we also observe workers exercising agency to improve their environments, sometimes without the involvement of design.

\subsection{Conflicts Across the Labor Chain}

Our first case illustrates how couriers' needs frequently collide with those of upstream and downstream actors, such as restaurant staff and consumers (see Figure \ref{multilaborer} in dark yellow). 
\revision{While food couriers are the most visible workers in HCI and design research, we found that restaurant workers and customers are equally entangled in the labor chain from a social interaction perspective~\cite{shuhao_centered_23}.}
Crucially, restaurant staff cannot be equated with ``capital'' or corporate owners. Most are low-wage employees in precarious jobs, with limited control over work conditions~\cite{Koo_serviceworker}.

This tension became evident during the 2025 Price War among three Chinese platforms. Aggressive discounts and zero-cost meal promotions drove order volumes in small restaurants and beverage stores to several times their usual capacity~\cite{pricewar_shanghaidaily}. 
Chefs, cooks, packers, and waiters, who are typically paid hourly or on fixed wages, shouldered the surge in workload without a corresponding increase in pay~\cite{pricewar_economist}. 
Reports of exhaustion and burnout from restaurant staff were widespread, as meal orders piled up faster than they could be prepared.
The couriers, meanwhile, were bound by strict delivery-time guarantees. At pickup points, kitchen staff often accused food couriers of rushing them, while food couriers in turn complained about delays. When meals inevitably arrived late, frustrated customers redirected blame to couriers. 
Consumers were also drawn into the conflict when the platforms introduced courier-friendly policies, such as canceling late-delivery fines or granting food couriers more flexibility. On social media, however, backlash quickly emerged, especially from white-collar workers reliant on punctual lunches during rigid breaks. In some contexts where speed and efficiency are deeply valued, even a 10–15 minute delay was framed as unacceptable. Ironically, policies intended to benefit couriers often left them more vulnerable to accusations of carelessness or unprofessionalism.

The platforms positioned themselves as neutral intermediaries, offering communication channels, while obscuring their own role in creating systemic tensions. This finding demonstrates that WCD interventions cannot be treated in isolation: improving one group's welfare inevitably reverberates across the labor chain, generating ripple effects for other workers as well as consumers.

\subsection{Distorted Worker-Centered Design}
\revision{Our interviews with platform designers reveal that some were genuinely motivated to improve courier welfare~\cite{shuhao_decide_24}}, often shaped by personal backgrounds (such as growing up in union families) or broader labor movements like U.S. gig worker strikes.
These worker-centered aspirations sometimes appeared in the form of worker-facing applications and at other times through shifts in business models. As shown in Figure \ref{multilaborer}, the light yellow segments illustrate how the platforms integrated food processing and warehousing to create their own retail channels and instant delivery services.

A prominent example was DoorDash's launch of DashMart in 2020, which expanded the company into logistics and warehouse, introduced a retail channel, and hired full-time couriers with benefits. \revision{In our study, designers view these initiatives as concrete efforts to stabilize working conditions and enhance security~\cite{shuhao_decide_24}.} Yet recent investigations painted a different picture: DashMart intensified managerial control and reduced flexibility~\cite{Sekharan_darkstore}. This contradiction exposes a paradox. When gig workers are offered benefits similar to employees, they often lose the autonomy once celebrated as flexibility, shifting precarity from gig work to more conventional workplace struggles, such as surveillance and constrained agency. WCD within platforms is therefore refracted through political and economic structures, producing partial gains that reconfigure rather than resolve fundamental labor issues.

The risk of distortion is not limited to platform-based initiatives. In one design probe (see Appendix \ref{appendix1}), we prototyped a crowd-sourcing platform that allowed couriers to share preferred waiting spots such as parks, shaded trees, or courier-friendly restaurants. Although intended to strengthen community ties, the system revealed hidden tensions in the wild. Many couriers resisted broad sharing, fearing the loss of ``territories'' to competitors. Others reappropriated the tool to flag ``difficult customers'' or ``unfriendly neighborhoods''. Although these adaptations generated useful peer knowledge, they also risked stigmatizing individuals and communities, producing harm beyond the scope of the design goals.

These findings show how difficult it is to design for workers. Even well-intentioned WCD can be reappropriated in ways that intensify competition, reproduce inequality, or shift precarity. Across different settings, WCD remains vulnerable to distortion, sometimes creating new problems while addressing old ones.

\subsection{Variations in Political-Economic Contexts and Designers' Unfamiliarity}
 
Over the four years we studied the food delivery sector, fast-changing political-economic contexts repeatedly reshaped both platform practices and our own conclusions.
Another tension revealed by the Multi-Laborer System is that while external political and economic factors significantly shape platforms, designers often lack awareness of these broader influences. This is particularly important for WCD research, as highlighted in Section \ref{relatedwork}, where much of the work places expectations on platform design. Designers and product managers are the ones who could potentially turn these opportunities into reality, yet they operate in environments that may not fully respond to broader structural forces~\cite{shuhao_decide_24}.

Designers' engagement with political-economic dynamics was limited.
In the European Union, regulations such as the General Data Protection Regulation (GDPR)~\cite{GDPR16} and the European Accessibility Act (EAA)~\cite{EUR_EAA19} imposed strict requirements on data transparency and accessibility. These compelled platforms to adjust their services to remain compliant. \revision{We found that platforms often pursued minimal compliance, tasking designers with rapid, formal changes that met legal requirements without advancing accountability, justice, or worker rights~\cite{shuhao_decide_24}}. Accessibility, though familiar to designers, was often deprioritized when it offered limited profit.

In addition to having limited policy awareness, another example is that designers have yet to fully recognize how their work influences consumer market and labor market dynamics. \revision{In our study, designers attributed platforms' neglect of worker needs to the abundance of labor~\cite{shuhao_decide_24, Ma_chi25}}: Europe's vulnerable migrant workforce and China's inexpensive labor pool ensured a steady supply of couriers, reducing corporate incentives to improve conditions. Designers, therefore, felt unable to advocate for change within corporates, especially when their own job security was uncertain. Yet business competition complicates this explanation. In 2025, the e-commerce company JD entered China's food delivery industry, which had long been dominated by Meituan and Eleme~\cite{pricewar_reuter2}. To attract couriers, JD introduced social welfare benefits, forcing competitors to follow suit~\cite{pricewar_chinadaily2}. Welfare thus became both a recruitment strategy and a way to project a benevolent corporate image. Despite uneven implementation, these measures expanded access to formal protections and improved the institutional environment for workers.

These cases also reshaped our understanding. Designers' limited engagement with political-economic changes is unsurprising: such issues are rarely taught in design or engineering education, and expecting designers to navigate them alone is unrealistic~\cite{Davis2023-va}. As tech workers embedded in corporate hierarchies, designers also face structural constraints that limit their ability to challenge market or political pressures, even when they care about workers~\cite{Su_techlas}.
Together, these cases show how volatile political-economic contexts, combined with designers' limited familiarity with them, constrain efforts to leverage policy or market shifts for worker welfare. WCD is inevitably shaped by these forces, yet designers remain poorly positioned to harness them.
The distortions of WCD described earlier are not isolated accidents but outcomes of a broader, rapidly shifting landscape. Changing regulations, fluctuating labor supplies, and sudden market competitions continually redefine the conditions under which designers and workers operate. To understand why WCD is so easily refracted or undermined, we must situate it within these wider dynamics.

\subsection{Workers as Agents of Change}

Another finding highlights that couriers, though often portrayed as precarious and passive, actively reshape their environments through everyday practices. This insight emerges from examining the outer boundaries of the food delivery Multi-Laborer System, which is not isolated but deeply entangled with other Multi-Laborer Systems, such as shopping malls and civic authorities. Over three years of observation in Lisbon, we documented how food couriers' repeated actions gradually transformed restaurants, public spaces, and even aspects of urban infrastructure~\cite{shuhao_centered_23}.

At fast-food chains such as McDonald's, couriers initially clustered outside, waiting in informal groups until staff handed over meals. Over time, restaurants began making accommodations: first by offering a spare table, then creating a designated waiting corner, and eventually installing dedicated pick-up counters with order display screens (see Figure \ref{photoAB} in Appendix). These changes shifted the interaction between food couriers and staff from ad-hoc negotiations to standardized workflows, demonstrating how food couriers' persistent presence prompted spatial and technological adjustments.
Similar dynamics unfolded in the urban landscape. Couriers, who are mainly dependent on bicycles and motorbikes, often struggle with inadequate parking facilities. At one major shopping mall, food couriers initially gathered inside the lobby or loitered near benches while precariously parking their vehicles wherever space allowed. Over time, couriers informally organized around the trees outside, leaning their bikes against trunks in orderly rows. As the number of food couriers grew, this informal practice evolved into a de facto parking system, eventually leading the mall and city authorities to introduce official motorcycle parking zones nearby (see Figure \ref{photoCD} in Appendix). What began as a workaround gradually reshaped both commercial and municipal infrastructure.
Worker agency is increasingly visible in Portugal through union–government negotiations that advocate for minimum and fairer wages, enabling workers to shape their conditions through unionization and policy awareness~\cite{Agencia_Lusa2025}. \revision{In contrast, our 2023 study found that migrant workers rarely commented on policy, as their primary concern was securing legal status in Portugal rather than engaging in labor advocacy~\cite{shuhao_centered_23}.}

Through these field studies and reflections, we observed that past WCD often revolved around platforms and systems, yet overlooked the fact that platforms alone cannot resolve the multiple identities and values entangled in workers' lives. 
In other words, WCD that fails to align with other interconnected Multi-Laborer Systems faces a high risk of failure.
At the same time, workers' active efforts to shape their own experiences resonate with the community-led principles of design justice~\cite{costanza-chock_design_2020}. This prompts us to reflect on whether design interventions should sometimes take a step back, first attending to the solutions that worker communities themselves are already developing.
Together, these examples highlight that couriers are not merely acted upon by platform design but actively reshape their environments, at times reinforcing, at times subverting the very systems in which they are embedded.

\section{A Diagnostic–Generative Pathway in Designing for Workers}
\label{pathway}

Building on our analysis, we identified tensions and blind spots in WCD projects. To make designing for workers more grounded, we adapt the double diamond model to articulate what we call the Diagnostic–Generative Pathway (Pathway). 
The Pathway is not meant for a single project, as it is unrealistic to expect one project to address every issue it raises. Instead, it is meant to reorient WCD and guide knowledge production across different stages toward more actionable outcomes. For researchers, our Pathway helps illuminate the complexity of workers' environments and navigate how design implications may or may not be adopted by platforms. For design practitioners, it offers a tool to support debate, negotiation, and decision-making in the corporate setting. For design educators, it invites reflection on the political and economic literacy needed when encouraging designers to tackle wicked problems.

Specifically, each phase is reoriented and reinterpreted. \textbf{\textit{Discover}} calls for attention to lived experiences while situating them within broader systems of interdependence. \textbf{\textit{Define}} frames needs not as isolated demands but as part of social and labor relations. \textbf{\textit{Develop}} emphasizes designing with awareness of potential distortions and conflicts, ensuring prototypes anticipate appropriation and unintended consequences. \textbf{\textit{Deliver}} highlights the importance of actionable designs that can withstand political-economic realities rather than remaining as utopian prototypes. This Pathway makes explicit the structural tensions and political-economic sensitivities that shape how design unfolds. In this section, we illustrate its use through examples.

\subsection{Problem–Solution: Conflict of Labor and Institutional Reframing}
Across the problem–solution trajectory (in red), our Pathway foregrounds two critical diagnostic nodes. 

The first is \textbf{\textit{Conflict of Labor}}, referring to the tensions that emerge when the needs of targeted workers intersect with or contradict those of other laborers in the system. As shown in Figure \ref{multilaborer}, prior studies often define worker needs in isolation. Our analysis shows that understanding these needs is insufficient, as they may be constrained or blocked by tensions within the broader Multi-Laborer System. Attending to how WCD is conditioned by other labor groups not only reshapes possible solutions but also opens opportunities for multi-laborer value co-creation and negotiation.
The second is \textbf{\textit{Institutional Reframing}}, which refers to the risk that WCD initiatives, whether developed by platforms, academia, or worker communities, are reframed to serve institutional interests (e.g., platforms, state actors, or worker organizations), reducing workers to means rather than ends.
Prior scholarship has highlighted how workers are often dehumanized, treated primarily as inputs to efficiency or productivity~\cite{Qadri_infras, Shaikh_chi24}.
We further draw inspiration from design under capitalism~\cite{wolf_descapitalism}, and from our analysis, we identify that both workers and designers can be co-opted by a capitalist world. This perspective helps us articulate how WCD interventions may likewise be reframed by diverse institutions and their stakeholders in practice.

In this line, two diagnostic points, conflict of labor and institutional reframing, define the conditions under which WCD takes shape.
Conflict of labor arises when the needs of focal workers intersect with or contradict those of other laborers in the system, making it clear that no single group's needs can be addressed in isolation. 
Reframing occurs when designs intended to support workers are co-opted for other purposes, such as enhancing organizational branding or increasing customer retention, while workers' interests fade into the background.
Addressing these dynamics requires attention not only to the political–economic contexts that shape workers' lives, but also to the political–economic position of designers themselves. This means recognizing the limits of designers' jurisdiction and strategically considering the language and arguments they can use when negotiating workers' needs with decision-makers.
Recognizing these dynamics opens up action spaces for designers and researchers. First, by recognizing conflicts in the labor chain, designers can frame problems in ways that anticipate negotiation and multi-laborer value co-creation, rather than privileging one group at the expense of another. 
Second, by anticipating how their designs may be reframed by institutions, designers can align proposals with institutional incentives to increase adoption while steering co-optation toward constructive outcomes, that is, using design reframing and appropriation for social good~\cite{noauthor_2016}. 
Together, these strategies allow designers to maintain a commitment to workers while addressing institutional priorities, opening space for dialogue, and enabling more resilient interventions.

\subsection{Landing–Iteration: Sensing and Imagining Policy and Economy}
Across the landing–iteration trajectory (in blue), the framework foregrounds two often underrepresented capabilities: policy imagination and economic imagination.

While the relationship between design, policy, and economics has been widely theorized~\cite{Murphy2015-kx, heskett_creating_2009, Tang_cscwlabor, yang_policy_24chi, Boztepe2024-tn, Bason_designforpolicy}, industry designers are neither policymakers nor economists. 
Although some design fields, such as service design, incorporate interdisciplinary knowledge and tools such as business modeling~\cite{Stickdorn18}, learning about policy and economics has not been formally integrated into the education or professional culture of designers.
In addition, in worker-centered research, there is a persistent tendency to frame design primarily directed toward social good, often distancing it from the economic and political infrastructures in which it is embedded. 
Ignoring these forces does not shield design from them; it only narrows the decision space available to designers. 

\revision{Our survey} shows that almost all platform designers engage directly in frontline labor tasks to understand workers' needs, often as a mandatory practice~\cite{shuhao_decide_24}. In food delivery, for example, designers are occasionally required to work as couriers or customer service agents to identify design opportunities. However, this evidence indicates that WCD is deeply dependent on its political–economic environment and cannot be achieved only through design efforts. Designers working on wicked problems must also find ways to reshape their discursive power.

In the landing and iteration phases, designers' task becomes persuading decision-makers to prioritize the deployment of WCD. The corporate priority list is always profit-driven, and profit is closely linked to policy and economic dynamics. From a policy perspective, designers rarely treat regulation as a design resource. Consider the European Accessibility Act (EAA): designers identified differences in how couriers used mobile interfaces during daytime versus nighttime deliveries, yet this accessibility concern was sidelined by product managers in favor of new feature rollouts. 
If we look at the timeline of the EAA, proposed in 2015~\cite{EUR_EAA15}, approved in 2019~\cite{EUR_EAA19}, and scheduled to take effect in 2025~\cite{EUR_EAA19}, the regulatory trajectory was visible years in advance. Such policy trends create actionable spaces for design. By remaining attentive to relevant regulations in their domain, designers can frame proposals as risk-avoidance strategies, thereby raising their organizational priority.
From an economic perspective, designers are familiar with creating value through design~\cite{heskett_creating_2009}, but often lack economic imagination. Platforms constantly adjust their business models to consolidate their market position, and the active involvement of designers in this process could significantly improve workers' conditions. In markets with abundant labor, companies may not fear shortages, but competition can trigger workforce churn, and reputational crises, such as public exposure of exploitative practices, can directly affect stock prices. The designers' ability to anticipate these dynamics allows them to forecast the economic consequences of poor design choices, a perspective that organizations take seriously~\cite{Ekbia_polieco}.

This highlights a missing dimension in current practice: the limited imagination of designers of how policy and economic forces can shape, and even positively reinforce, their work. 
We argue instead that cultivating political and economic literacy should be treated as a core element of design education and practice, much like the adoption of psychological or ethnographic methods from the social sciences. 
Such literacy does not mean designers must become experts in economics or law. 
Rather than focusing only on the macro-level of political economy, we ask how designers can mobilize political and economic materials in support of their design claims. This does not require overambition. A pragmatic starting point is for designers to monitor and interpret relevant laws and regulations, using them both to test their designs and to frame proposals. 
Our Pathway enables them to anticipate how interventions may be reframed by institutions, and to strategically align with, or at times contest, policy and market dynamics in ways that expand the actionable space for justice-oriented design.
This awareness enables them to articulate worker-centered proposals in terms that resonate with decision-makers, increasing the likelihood of adoption. In this sense, policy and economic imagination are not only contextual sensitivities but also resources for value co-creation and for mitigating conflict among multiple stakeholders.
\section{Discussion}
We conclude by discussing needed transitions in designing for workers, the broader relevance of our study across domains, its limitations, and directions for future research.

\subsection{Challenges in Worker-Centered Design}
Interest in social justice has been steadily growing within HCI and design~\cite{Chordia_justice_in_hci}, with designing for workers emerging as a central concern~\cite{fox_workercenter_dis20, chiwork_22}.
A range of methods, frameworks, and experiments have been proposed to address this complex challenge~\cite{Bardzell_feminist10, Friedman2019-VSD, Dombrowski_SJOD, penin_bloomsbury_2026}.
In outlining social-justice-oriented interaction design, \citet{Dombrowski_SJOD} emphasized commitments to conflict, reflexivity, ethics, and politics.
\citet{penin_bloomsbury_2026} developed worker perspectives within the service design tradition, highlighting that service work itself has been shaped by the practices of service design.

Despite WCD's promise, our four-lens framework reveals persistent tensions in implementing WCD in practice. 
From the first lens, WCD may overlook how design decisions shape broader labor systems, influenced by longstanding design limitations such as those found in UCD. Because WCD unfolds within a Multi-Laborer System, centering a single worker group is insufficient for addressing wicked problems like worker justice.
From the second lens, designing for workers in both corporate and academic settings is inherently complex. Well-intentioned efforts can be co-opted to intensify competition, inequality, or precarity, and our own prototype deployment showed that WCD can unintentionally create new challenges.
From the third lens, our work shows that some platform designers make sincere attempts to improve worker experiences, yet lasting impact remains limited. Their constrained economic-political imagination further limits the effectiveness of WCD. This suggests that platform designers rarely encounter or adopt WCD research, highlighting a new action space between WCD researchers and platform designers.
From the fourth lens, emerging competition, policies, and technologies continue to reshape work, introducing new business models, tensions, and opportunities that continually challenge WCD.
Across academic and industry settings, these challenges help us reflect on the action spaces of WCD and better understand the forms of real-world impact it seeks to achieve, including the distinct spaces where researchers, practitioners, and collaborators can act.
In the next section, we discuss two key transitions our work suggests across design disciplines and target audiences, including UX, product, and service design, whether consumer-facing, business-facing, or worker-facing.

\subsection{Transitions in Worker-Centered Design}

First, our findings reveal that WCD requires a significant methodological reflection. Even when researchers and designers aim to support worker welfare, their interventions may be co-opted by platform structures, reinforcing rather than reducing precarity. 
Researchers are increasingly cautioning that the design itself can become a tool of the systems they seek to challenge~\cite{Mareis2022-kx, Boehnert2014-dl, DiSalvo_12}. \citet{Boehnert2014-dl} exposes how the systemic and capitalist priorities of the design industry often reproduce unsustainable conditions, despite the intentions of designers.
Also, the simple inclusion of workers' voices in the research process does not resolve their problems, aligning with the focus of value-sensitive design on identifying and negotiating stakeholder conflicts and value tensions~\cite{Friedman2019-VSD}.
It also prompts renewed reflection on PD and co-design~\cite{Moll2020-pk, Wacnik2025-cm}, urging scrutiny of whether current forms of worker participation meaningfully address the systemic conditions shaping work.

The complexity of conflicts, the limits of designers, and the volatility of political-economic contexts make it difficult to translate commitments into practice. 
We embed this reflexivity into its design process, foregrounding labor conflicts and the risk of institutional reframing as diagnostic nodes that designers must consciously navigate. 
We recommend that WCD researchers draw on multi-laborer conflicts and the Pathway when proposing new opportunities. This perspective helps clarify what is required for WCD to take effect in practice and encourages reflection on whether interventions might be co-opted through institutional reframing or produce unintended risks. By examining these tensions and blind spots, the Pathway supports laying a foundation for new action spaces such as value co-creation~\cite{Aarikka-Stenroos2012-un, Perna2022-ac}.

The second transition lies in the language and rhetoric used by designers~\cite{Adams2022-vl, Bucciarelli94, krippendorff_semantic_2006}. 
WCD must engage with policy, organizational strategy, and corporate interests, navigating diverse and often conflicting priorities among institutions.
Within academic and corporate contexts, designing for worker framings may alienate consumer-focused or business-oriented stakeholders. 
\citet{Mulligan_callfairness} highlighted the importance of shared language grounded in inclusivity and fairness, yet they also note that translation across stakeholder groups remains deeply challenging. 
We emphasize the importance of incorporating diverse worker perspectives and analyzing labor conflicts as rhetorical resources that can strengthen designers' positions in negotiations with corporate stakeholders. 
It reflects a design tradition often overlooked in WCD research, where conflicts serve not only as barriers but also as creative entry points to envision alternative solutions~\cite{Dorst2001-kv, krippendorff_semantic_2006}.
This transition calls for broader adoption and opens opportunities for collaboration and mutual recognition across political-economic contexts.

\subsection{Broader Applications of Multi-Laborer System and Pathway}
Our four‑lens analysis of the food delivery industry highlights the concept of a Multi-Laborer System and introduces the Pathway. As explained in Section \ref{method}, we derived these two concepts through reflexive analysis, informed by our background in service design, which helped us visually map the tensions.

We conceptualize the Multi-Laborer System as a tool to visualize diverse labor roles and hidden tensions within service platforms. It refers to a labor chain where different types of workers, each with distinct roles, collectively support the operation and delivery of a service ecology.
In this system, designers are also laborers constrained by interdisciplinary teams and conflicting stakeholders' interests. The tool helps them identify labor roles and actionable gaps, revealing systemic tensions and new opportunities for value co-creation. Crucially, we argue that both workers and consumers are laboring subjects, and framing consumers only as privileged users may limit platform adoption of WCD. 
This view aligns with critiques of platform and surveillance capitalism~\cite{Srnicek2016, Zuboff2019}, where all users contribute labor and are subject to platform control.
The Pathway builds on the Multi-Laborer System as a framework for envisioning the future of WCD. It is not a linear design process, and its efforts may engage with one or several parts of it. Advancing this framework requires the collective efforts of researchers, design practitioners, and educators to meaningfully address the wicked problems of worker justice.
More specifically, we highlight three directions for future application.

First, within design teams. Design practitioners often represent distinct stakeholder groups such as consumers, couriers, or merchants. These teams may work in parallel while pursuing goals that are not only different but sometimes in conflict. A multi-laborer perspective creates opportunities to move from simply informing, such as revealing how the needs of one group affect others, to coordinating across teams in order to reduce predictable tensions. Echoing service research~\cite{Hakanen2012, Oertzen2018}, when conditions allow, this can lead to co-creating solutions that aim to balance competing interests. This process helps ease friction between teams and contributes to more sustainable and equitable design outcomes.

Second, across industries. While our case focuses on food delivery, similar Multi-Laborer Systems exist in ride-hailing (e.g., fleet, drivers, and passengers), healthcare (e.g., clinicians, nurses, care workers, and patients), and other service sectors where diverse workers, clients, and institutions interact~\cite{penin_bloomsbury_2026}. Meanwhile, the Pathway helps surface hidden conflicts and reveals how design may be used in ways that contradict its original intent. By making these dynamics visible, the approach supports more reflective and accountable design decision-making across different contexts.

Third, across design artifacts. We hope our work speaks not only to those in interaction and user experience design, but also to practitioners engaged in service design, architectural design, and organizational design. These artifacts are shaped by distinct yet interconnected labor systems, but they often portray workers as static roles rather than as participants in evolving labor relations. 
For example, architecture is a labor-intensive industry. When architectural and spatial design reflect on how their practices influence tensions within Multi-Laborer Systems (e.g., constructors), they can contribute to more inclusive transitions~\cite{ursprung_representacion_2018, deamer2020_archi_labor}.
Bringing labor dynamics into the design of both physical and digital artifacts can deepen awareness of the social nature of design~\cite{krippendorff_semantic_2006, Dantec_Situated}, better ground it in the lived conditions of work, and support more thoughtful consideration of our effects on the working class.

\subsection{Limitations}
The proposed Multi-Laborer System and the Diagnostic–Generative Pathway emphasize reflexivity and generativity, but whether designers have the time or authority to apply them in practice remains uncertain. Understanding these adoption barriers is an important direction for future work. One possibility is to invite designers from different industries to reflect on their past projects through the Pathway.
We also recognize that our work may raise criticism. Some researchers may argue that justice requires prioritizing marginalized groups rather than balancing across stakeholders. We do not challenge such perspectives or any specific politics of method. Our intention is instead to offer inspiration about actionable spaces for designers and researchers, particularly those who are beginning to engage with worker-centered questions.
Last, advancing worker justice and well-being is a long-term endeavor. Our Pathway does not offer direct solutions but aims to reorient designers' mindsets. We encourage further research and education to build on this orientation and provide stronger foundations for designing justice in practice.
\section{Conclusion}
In this paper, we provide a critical revisit of WCD using a four-lens framework comprising foundation, frontstage, backstage, and offstage. This approach reveals tensions, blind spots, and systemic constraints that limit the effectiveness of current WCD efforts.
From this analysis, we introduce two key concepts: the Multi‑Laborer System, which offers a holistic view of labor relationships and conflicts, and the Diagnostic–Generative Pathway, which helps open action spaces for design interventions that take political, economic, and organizational realities into account.
Our contribution is threefold. First, we reflect on the gap between WCD research and its intended real‑world impact. Second, we offer a model that captures the complexity of labor systems and highlights where design must intervene. Third, we sketch a practical orientation to guide researchers and designers toward more just, inclusive, and practical outcomes. We hope this work advances the bridge between research and practice and invites continued collaboration, critical reflection, and education to expand the potential of design for workers.

\begin{acks}
This work is co-financed by Fundação para a Ciência e a Tecnologia (FCT) through the Carnegie Mellon Portugal Program under the fellowship FCT Ph.D. Grant PD/BD/152201/2021.
The research is funded by ITI/LARSyS through FCT funding, under projects 10.54499/LA/P/0083/2020 and UID/50009/2025. We thank the workers, designers, researchers, and design educators who shared their time and insights with us over the past four years.
\end{acks}
\bibliographystyle{ACM-Reference-Format}
\bibliography{References/CHI2025_Work_Vision}

\appendix

\newpage
\onecolumn

\section{A Design Probe} \label{appendix1}
The appendix A presents the design probe prototype.

\begin{figure}[h]
  \centering
  \includegraphics[width=\textwidth]{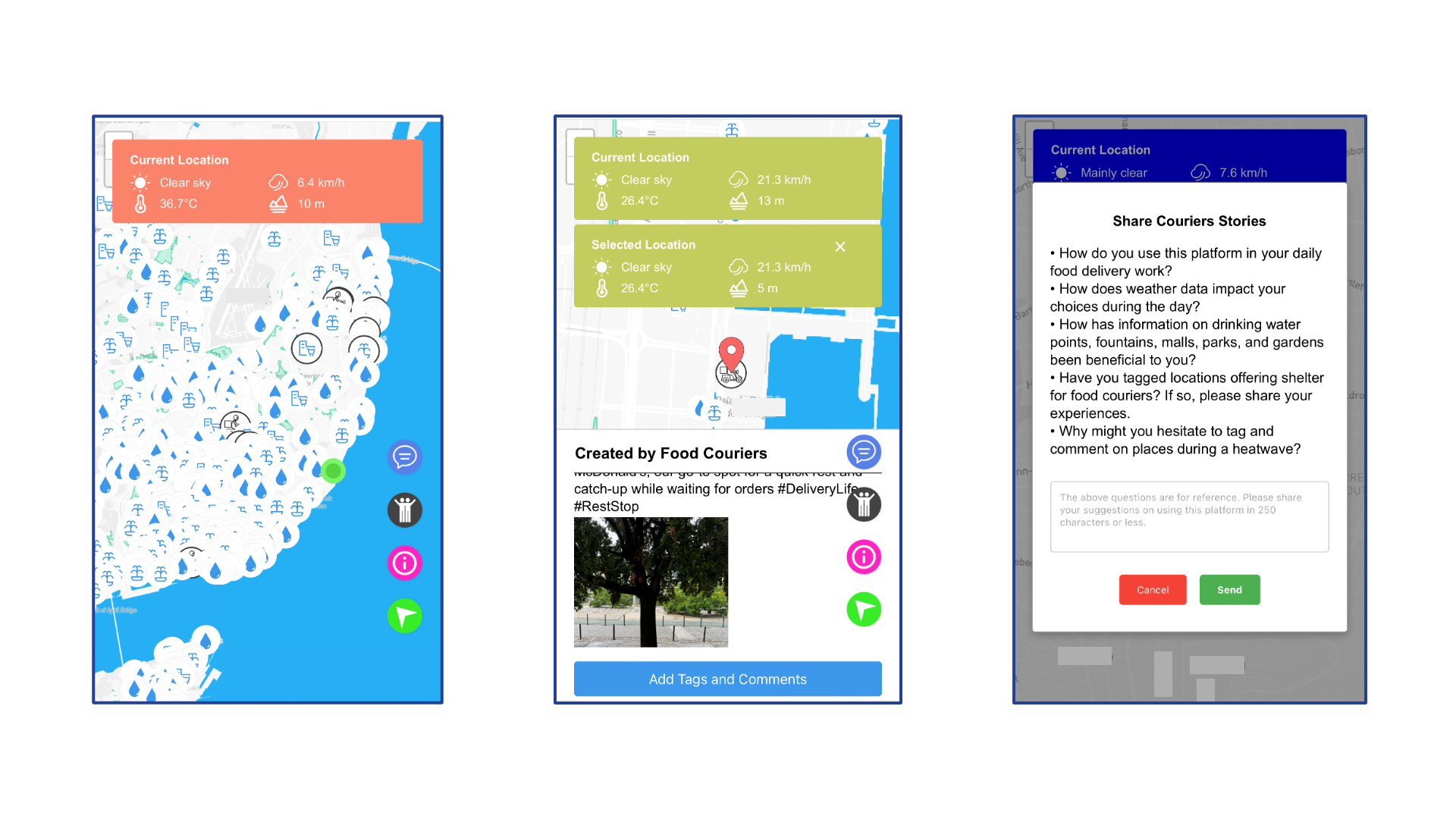}
  \caption{\textbf{A Design Probe:} This figure shows a prototype of a worker-centered, crowdsourced platform for food couriers. The platform integrates open environmental data (e.g., weather, drinking water points, fountains, malls; left) and enables couriers to tag meaningful locations and share experiences (middle and right). Designed as a technology probe, it aims to surface how couriers creatively repurpose urban spaces and develop community practices around them.}
  \Description{This figure shows a prototype of a worker-centered, crowdsourced platform for food couriers. The platform integrates open environmental data (e.g., weather, drinking water points, fountains, malls; left) and enables couriers to tag meaningful locations and share experiences (middle and right). Designed as a technology probe, it aims to surface how couriers creatively repurpose urban spaces and develop community practices around them.}
  \label{designprobe}
\end{figure}

\clearpage

\section{Fieldwork Photos} \label{appendix2}

This appendix presents three sets of photos taken during fieldwork over the past four years, serving as supplementary documentation for the results. These images are included here because the analysis and presentation of results remain clearly understandable without them.

\vspace{1\baselineskip}

\begin{figure}[h]
  \centering
  \includegraphics[width=\textwidth]{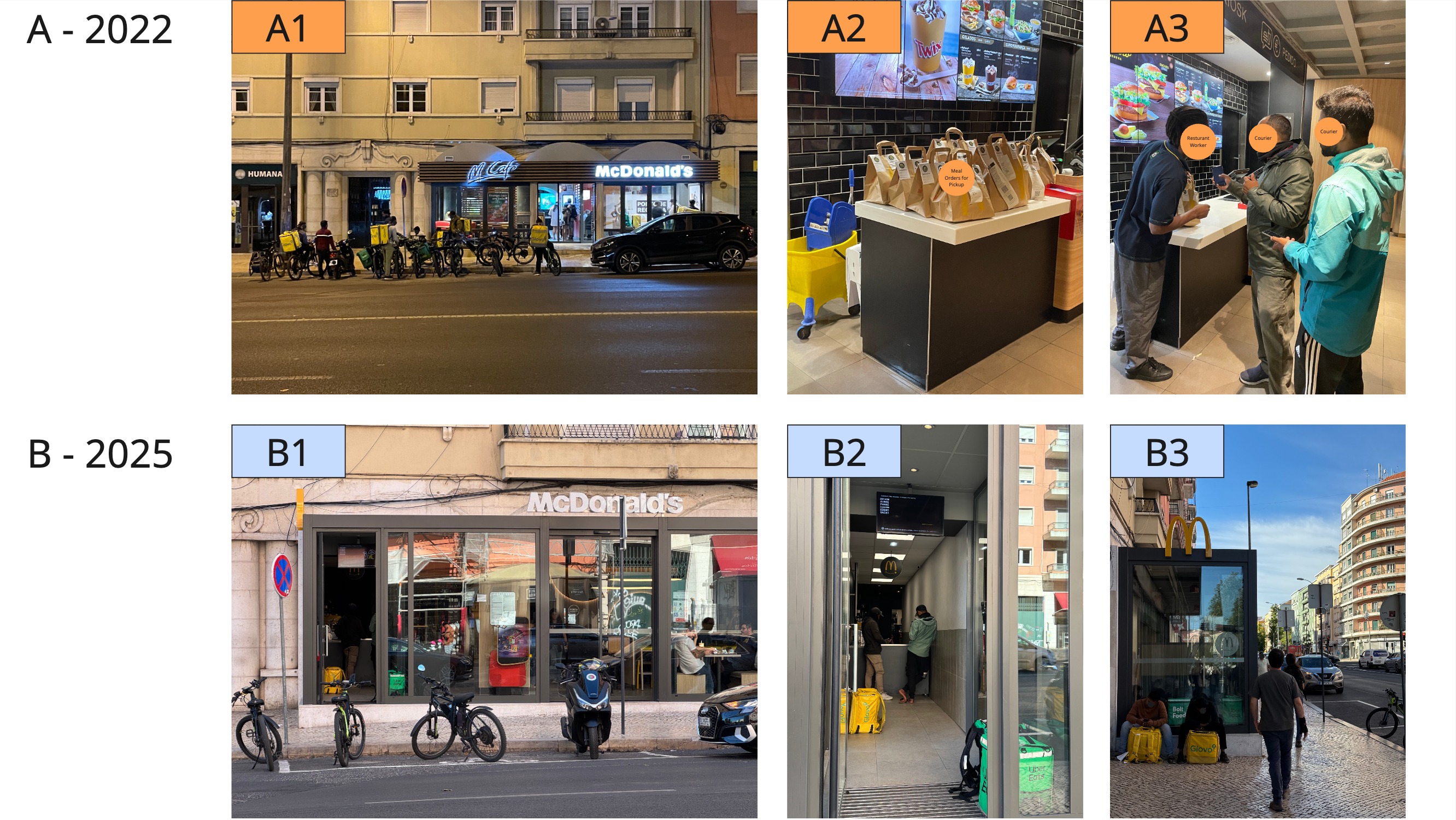}
  \caption{\textbf{Observation on a McDonald Shop:} This set of photos documents changes in delivery service touchpoints over three years at the same McDonald’s in a European capital. Set A shows the typical interaction: couriers wait at the entrance, the kitchen prepares meals, staff package them, and the handoff occurs through verbal order confirmation. Set B highlights changes driven by digital technologies. As delivery volume grew, couriers increasingly entered the restaurant, overlapping with dine-in customers. Then, the restaurant created a dedicated pickup corridor on the left side, with a dashboard displaying order status, enabling contactless retrieval.}
  \Description{This set of photos documents changes in delivery service touchpoints over three years at the same McDonald’s in a European capital. Set A shows the typical interaction: couriers wait at the entrance, the kitchen prepares meals, staff package them, and the handoff occurs through verbal order confirmation. Set B highlights changes driven by digital technologies. As delivery volume grew, couriers increasingly entered the restaurant, overlapping with dine-in customers. Then, the restaurant created a dedicated pickup corridor on the left side, with a dashboard displaying order status, enabling contactless retrieval.}
  \label{photoAB}
\end{figure}

\begin{figure}[h]
  \centering
  \includegraphics[width=\textwidth]{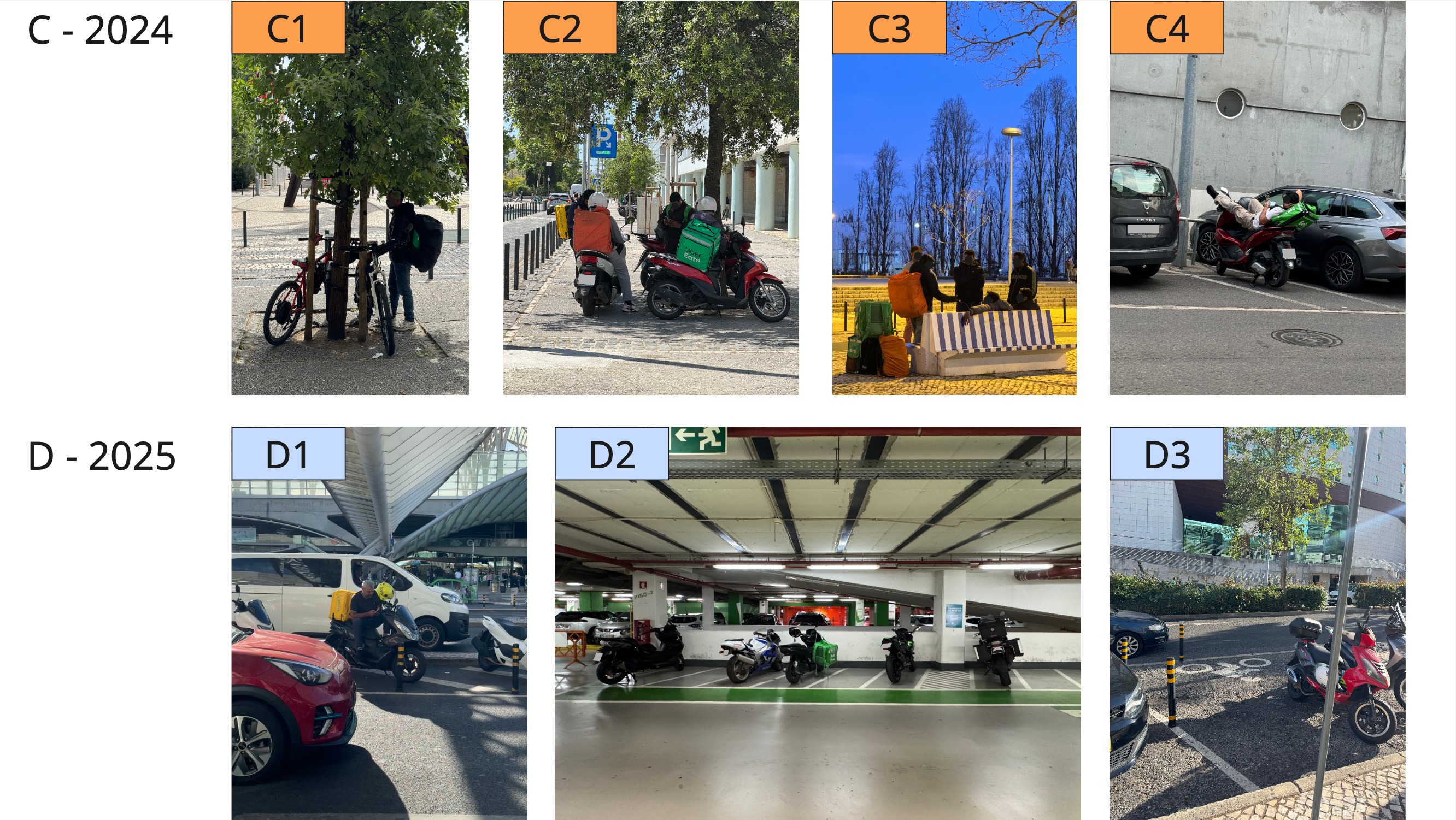}
  \caption{\textbf{Observation on Parking Spaces nearby a Shopping Mall:} This set of photos documents how couriers park around a large shopping mall. Set C, taken in 2024, shows that neither the mall nor nearby public infrastructure was prepared for the surge in motorcycles and bicycles. As a result, couriers informally appropriated public space — for example, locking bikes to trees (C1) or using gaps between car parking spots for motorcycle parking (C4). Set D shows changes in 2025, when dedicated motorcycle parking areas were introduced around the mall and in its underground garage.}
  \Description{This set of photos documents how couriers park around a large shopping mall. Set C, taken in 2024, shows that neither the mall nor nearby public infrastructure was prepared for the surge in motorcycles and bicycles. As a result, couriers informally appropriated public space — for example, locking bikes to trees (C1) or using gaps between car parking spots for motorcycle parking (C4). Set D shows changes in 2025, when dedicated motorcycle parking areas were introduced around the mall and in its underground garage.}
  \label{photoCD}
  \addvspace{2em}
\end{figure}

\begin{figure}[h]
  \centering
  \includegraphics[width=\textwidth]{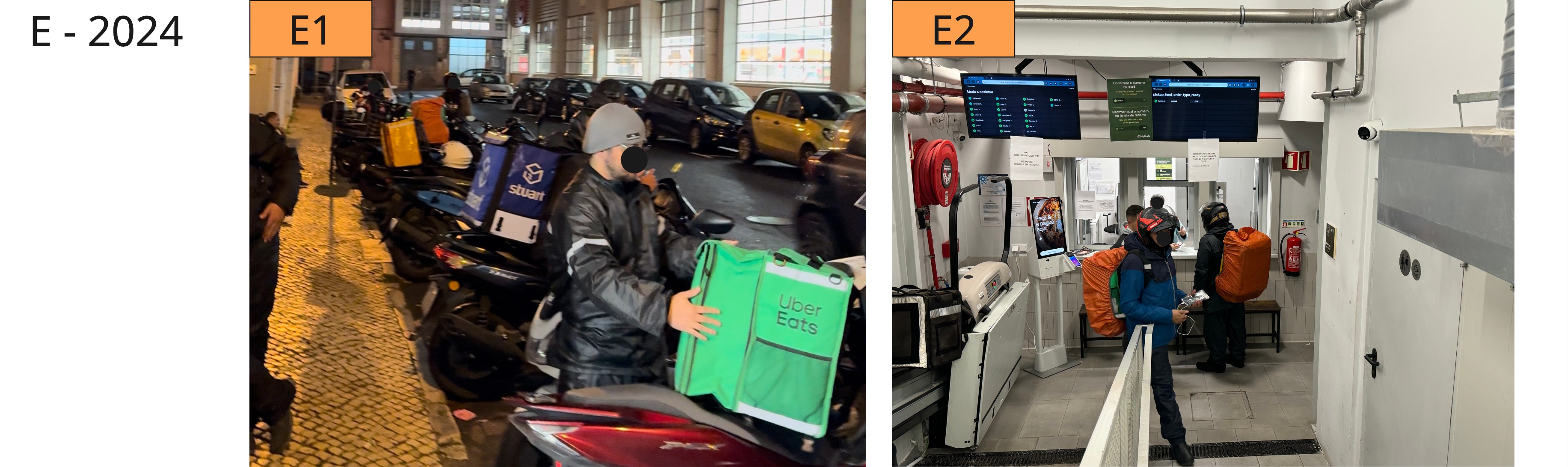}
  \caption{\textbf{Observation on a Central Kitchen:} This image depicts both the exterior and interior of a central kitchen, which prepares meals for multiple restaurants. It does not offer dine-in services and is exclusively dedicated to delivery. The kitchen space itself is not visible to the public. Outside, motorcycles and bicycles belonging to couriers are almost always lined up at the entrance. Inside, the space is organized around pickup windows, a dashboard, and a narrow corridor. Couriers simply wait for their assigned number to appear on the dashboard, then retrieve their orders either from front-desk staff or directly from designated lockers. The pickup process can be entirely contactless, requiring no verbal interaction.}
  \Description{This image depicts both the exterior and interior of a central kitchen, which prepares meals for multiple restaurants. It does not offer dine-in services and is exclusively dedicated to delivery. The kitchen space itself is not visible to the public. Outside, motorcycles and bicycles belonging to couriers are almost always lined up at the entrance. Inside, the space is organized around pickup windows, a dashboard, and a narrow corridor. Couriers simply wait for their assigned number to appear on the dashboard, then retrieve their orders either from front-desk staff or directly from designated lockers. The pickup process can be entirely contactless, requiring no verbal interaction.}
  \label{photoE}
\end{figure}

\end{document}